\documentclass[ aps, showpacs, showkeys, nofootinbib, floatfix, superscriptaddress]{revtex4}

\usepackage{amsfonts}

\usepackage{amssymb}
\usepackage{amsmath}
\usepackage{graphicx}
\usepackage{hyperref}
\begin{document}

\title{Time variable cosmological constant  of
 holographic origin  with interaction in Brans-Dicke theory}

 \author{Jianbo Lu}
 \email{lvjianbo819@163.com}
 \affiliation{Department of Physics, Liaoning Normal University, Dalian 116029, P. R. China}
 \author{Lina Ma}
 \affiliation{Department of Physics, Liaoning Normal University, Dalian 116029, P. R. China}
 \author{Molin Liu}
 \affiliation{College of Physics and Electronic Engineering, Xinyang Normal University, Xinyang, 464000, P. R. China}
 \author{Yabo Wu}
 \affiliation{Department of Physics, Liaoning Normal University, Dalian 116029, P. R. China}

\begin{abstract}
 Time variable cosmological constant (TVCC) of holographic origin  with
interaction  in Brans-Dicke theory is discussed in this paper. We
investigate some characters for this model, and show  the evolutions
of deceleration parameter and equation of state (EOS) for dark
energy. It is shown that in this scenario an accelerating universe
can be obtained and the evolution of EOS for dark energy can cross
over the boundary of phantom divide. In addition, a geometrical
diagnostic method, jerk parameter is applied to this model to
distinguish it with cosmological constant.

\end{abstract}


\maketitle

\section{$\text{Introduction}$}

{The astronomical observations \cite{SNIa,SNIa1} support that the
universe is undergoing accelerated expansion. It indicates that
there exist a new component dubbed as dark energy with negative
pressure in universe. A natural explanation on the accelerated
expansion is due to a positive tiny cosmological constant of
holographic origin. Though the cosmological constant suffers the
so-called fine tuning and cosmic coincidence problems, it fits the
observations very well in 2$\sigma$ confidence level
\cite{LCDM2sigma,LCDM2sigma1}. However, if the cosmological constant
is not a constant but a time variable one
\cite{TVCC,TVCC1,TVCC2,quintom,DEmodels,DEmodels1,DEmodels2,DEmodels3,DEmodels4,DEmodels5,DEmodels6,DEmodels7,DEmodels8}
or there is an interaction between dark matter and dark energy
\cite{interaction,interaction-a,interaction1,interaction1a,interaction1b,
interaction1c,interaction1d,interaction1f,interaction1g,interaction1h,interaction1i,interaction1j},
the fine tuning and cosmic coincidence problems can be removed. In
particular, the dynamic vacuum energy density based on holographic
principle is investigated extensively
\cite{holographic,holographic1,holographic2,holographic3,holographic4}.
According to the holographic principle, the total energy of a system
with size $L$ should not exceed the mass of a black hole with the
same size as the system, and the largest $L$ allows a energy density
$\rho_{\Lambda}=3c^{2}M_{p}^{2}L^{-2}$, where $c$ is a numerical
constant and $M_{p}$ is the reduced Planck Mass $M_{p}^{-2}=8\pi G$.
It just means a duality between UV cut-off and IR cut-off which are
related to the vacuum energy and the large scale of the universe,
respectively. For example, Refs. \cite{8-12HDE,8-12HDE1} have
applied this principle to discuss Hubble horizon, future event
horizon and particle horizon in cosmology. Another dark energy model
with relations with holographic dark energy, named agegraphic dark
energy, was also extensively researched
\cite{19age,19-21age,19-21age1,19-21age2,19-21age3,19-21age4}. This
model is based on the
 well-known Heisenberg uncertainty relation to the
universe. Therefore, the energy density of metric fluctuations in
Minkowski space-time is $\rho_{\Lambda}\sim M_{p}^{2}/t^{2}$, where
$t$ is time or length scale. Obviously, it looks like the
holographic one, and there are some relations between them
\cite{19age}.

 In addition,  it is known that the de Sitter spacetime is described by the line element,
\begin{equation}
ds^{2}=(1-\frac{r^{2}}{l^{2}})dt^{2}+(1-\frac{r^{2}}{l^{2}})^{-1}dr^{2}+r^{2}(d\theta^{2}+sin^{2}\theta
d \varphi^{2}),
\end{equation}
where the cosmological constant is taken as $\Lambda=3/l^{2}$.
According to the horizon equation,  cosmological horizon is
obtained,
\begin{equation}
r_{\Lambda}=c^{'}t_{\Lambda}=\sqrt{3/\mid \Lambda \mid},
\end{equation}
where $c^{'}$ is the speed of light taken to unit here. Conversely,
any cosmological length scale or time scale can introduce a energy
density $\Lambda(t)$   into Einstein's theory
\begin{equation}
\Lambda(t)=\frac{3}{r^{2}_{\Lambda}(t)}=\frac{3}{t^{2}_{\Lambda}(t)}.
\end{equation}
When a dynamic time scale is taken into account, a time variable
cosmological constant (TVCC) is obtained. And a natural time scale
is the age of our universe,
$t_{\Lambda}=\int_{0}^{t}dt^{'}=\int_{0}^{a}\frac{da^{'}}{a^{'}H}$.
If  ones take this cosmological horizon as the length scale (or time
scale) of cut-off in de Sitter universe,
 vacuum energy density is
expressed, $ \rho_{\Lambda}=3c^{2}M_{p}^{2}/t_{\Lambda}^{2}$
\cite{Xu-plb}.
 From this point,, one can see that this time variable cosmological
constant  is similar with the holographic  and the agegraphic dark
energy.

On the other hand, Brans-Dicke theory (BD) \cite{21BD} as a natural
extension to Einstein's gravitational  theory of general relativity
(GR), can pass the experimental tests from the solar system
\cite{22BDsolar}. In Brans-Dicke theory, the gravitational
"constant" $G$ is replaced with the inverse of a time-dependent
scalar field $\Phi (t)$, which couples to gravity with a coupling
parameter $\omega$. The holographic dark energy model in the
framework of Brans-Dicke
 theory has already been considered
 by many people \cite{28-32HDEBD,28-32HDEBD1,28-32HDEBD2,28-32HDEBD3,28-32HDEBD4}.  Here
 we study some characters of time variable cosmological constant in BD
 theory.

\section{$\text{Interacting time variable cosmological constant in Brans-Dicke theory}$}

In the framework of  BD theory, the field is described by the action
below,
\begin{equation}
S=d^{4}x\sqrt{-g}[\Phi R
-\frac{\omega}{\Phi}\Phi^{,\alpha}\Phi_{,\alpha}+L_{m}],
\end{equation}
where $L_{m}$ is the matter Lagrangian, $\Phi$ is the  non-minimally
coupled BD scalar field which plays the role of
  gravitational "constant" $G$ with relating to the inverse of
the $\Phi$. The BD coupling parameter $\omega$ is restricted   to be
around 40000 according to the local gravity tests in the solar
system \cite{1005.0868-6}.  Some proposals address that when the
system changes from local to large cosmological scales  $\omega$
should have more smaller value
\cite{w-1000,0510779-14-15,0510779-14-15-a}.
 By varying above action with respect to
the metric, the gravitational field equations is given as
\begin{equation}
 G_{\mu\nu}=\frac{1}{\Phi}T^{m}_{\mu\nu}+\frac{\omega}{\Phi^{2}}
 [\Phi_{,\mu}\Phi_{,\nu}-\frac{1}{2}g_{\mu\nu}\Phi_{,\alpha}\Phi^{,\alpha}]
 +\frac{1}{\Phi}[\Phi_{,\mu;\nu}-g_{\mu\nu}\frac{
 T}{3+2\omega}],
 \end{equation}
where $G_{\mu \nu}$ is Einstein tensor,  $T_{\mu \nu}^{m}$ denotes
the energy-momentum tensor of matter, and
$T=T_{\mu\nu}^{m}g^{\mu\nu}$. If we consider the detailed components
of universe including the baryon matter $\rho_{b}$, the cold dark
matter $\rho_{c}$, the radiation $\rho_{r}$, and the time variable
cosmological constant $\rho_{\Lambda}$ as dark energy, the
gravitational equations are written as
\begin{equation}
3\Phi[H^{2}+H\frac{\dot{\Phi}}{\Phi}-\frac{\omega}{6}\frac{\dot{\Phi}^{2}}{\Phi^{2}}]=\rho,\label{Friedmann}
\end{equation}
\begin{equation}
2\frac{\ddot{a}}{a}+H^{2}+\frac{\omega}{2}\frac{\dot{\Phi}^{2}}{\Phi^{2}}+2H\frac{\dot{\Phi}}{\Phi}
+\frac{\ddot{\Phi}}{\Phi}=-\frac{p_{\Lambda}}{\Phi},\label{accelerating}
\end{equation}
where $H=\frac{\dot{a}}{a}$ is the Hubble parameter,
$\rho=\rho_{b}+\rho_{c}+\rho_{r}+\rho_{\Lambda}$,
 $p_{\Lambda}$
is the pressure of TVCC, and its energy density  is
\begin{equation}
\rho_{\Lambda}=3c^{2}\Phi(t)/t_{\Lambda}^{2},\label{tL}
\end{equation}
 with time scale
\begin{equation}
t_{\Lambda}=\int^{t}_{0}dt^{'}=\int^{a}_{0}\frac{da^{'}}{a^{'}H^{'}}.\label{rhotL}
\end{equation}
 Submitting
Eq. (\ref{rhotL}) into (\ref{tL}) and taking the derivative with
respect to $\ln a$, we get
\begin{equation}
\rho_{\Lambda}^{'}=\frac{d\rho_{\Lambda}}{d\ln a}=-\frac{6}{c}\Phi
H^{2}\Omega_{\Lambda}^{\frac{3}{2}}+3\Phi^{'}H^{2}\Omega_{\Lambda},\label{rhoPl}
\end{equation}
with the definition of dimensionless energy density,
$\Omega_{i}=\frac{\rho_{i}}{3H^{2}\Phi}$, where  $i$ can denote
radiation $\Omega_{r}$, baryon matter $\Omega_{b}$, cold dark matter
$\Omega_{c}$, and TVCC $\Omega_{\Lambda}$, respectively. We consider
an interaction existed in the two dark components, the energy
conservation equation of each component in universe is written
respectively as,
\begin{equation}
\dot{\rho}_{r}+4H\rho_{r}=0,
\end{equation}
\begin{equation}
\dot{\rho}_{b}+3H\rho_{b}=0,
\end{equation}
\begin{equation}
\dot{\rho}_{c}+3H\rho_{c}=Q,
\end{equation}
\begin{equation}
\dot{\rho}_{\Lambda}+3H(1+w_{\Lambda})\rho_{\Lambda}=-Q.\label{IL}
\end{equation}
By adopting  a generalized interacting term,
$Q=\lambda_{1}H\rho_{c}+\lambda_{2}H\rho_{\Lambda}$, Eq.(\ref{IL})
is rewritten in a form as
\begin{equation}
\rho_{\Lambda}^{'}+3(1+w_{\Lambda})\rho_{\Lambda}=-(\lambda_{1}\rho_{c}+\lambda_{2}\rho_{\Lambda}).\label{rhoP2}
\end{equation}
 Combining
Eqs.(\ref{rhoPl}) and (\ref{rhoP2}), we obtain the equation of state
(EOS) of this interacting time variable cosmological constant as,
\begin{equation}
w_{\Lambda}=\frac{2}{3c}\Omega_{\Lambda}^{\frac{1}{2}}-\frac{1}{3}
\frac{\Phi^{'}}{\Phi}-\frac{\lambda_{1}\Omega_{c}}{3\Omega_{\Lambda}}-\frac{\lambda_{2}}{3}-1.\label{w}
\end{equation}
With considering a parameterized
 form for scalar field $\Phi=\Phi_{0}(\frac{a}{a_{0}})^{\alpha}$,
  Friedmann equation (\ref{Friedmann}) is rewritten as
\begin{equation}
H^{2}=\frac{2 \rho}{(6+6\alpha-\omega \alpha^{2})\Phi}.
\label{friedmann2}
\end{equation}
Using Eqs. (\ref{Friedmann}), (\ref{accelerating}}) and
(\ref{friedmann2}), we get
\begin{equation}
\dot{H}=\frac{-3}{(6+6\alpha-\omega
\alpha^{2})\Phi}[(\rho+p_{\Lambda})+\frac{\alpha}{3}\rho].\label{Hdot}
\end{equation}
 Substituting Eq.(\ref{w}) into (\ref{rhoP2})
and applying the definition of $\Omega_{\Lambda}$, we obtain
\begin{equation}
\frac{H^{'}}{H}=\frac{1}{2\Omega_{\Lambda}}[-\Omega_{\Lambda}^{'}-\frac{2}{c}\Omega_{\Lambda}^{\frac{3}{2}}].\label{Hp1}
\end{equation}
On the other hand, substituting $\dot{H}=H^{'}H$ and
$p_{\Lambda}=w_{\Lambda}\rho_{\Lambda}$ into Eq.(\ref{Hdot}), we
have
\begin{equation}
\frac{H^{'}}{H}=\frac{-9}{6+6\alpha-\omega\alpha^{2}}[\Omega_{r}+\Omega_{m}+\frac{2}{3c}\Omega_{\Lambda}^{\frac{3}{2}}
-\frac{1}{3}\frac{\Phi^{'}}{\Phi}\Omega_{\Lambda}
-\frac{\lambda_{1}}{3}\Omega_{c}-\frac{\lambda_{2}}{3}\Omega_{\Lambda}+\frac{\alpha}{3}(\Omega_{r}+\Omega_{m}+\Omega_{\Lambda})],\label{Hp2}
\end{equation}
with $\Omega_{m}=\Omega_{b}+\Omega_{c}$. Thus, combining
Eqs.(\ref{Hp1}) and (\ref{Hp2}) we get the differential equation for
$\Omega_{\Lambda}$,
\begin{equation}
\frac{\Omega_{\Lambda}^{'}}{\Omega_{\Lambda}}=\frac{18}{6+6\alpha-\omega
\alpha^{2}}[\Omega_{r}+\Omega_{m}+\frac{2}{3c}\Omega_{\Lambda}^{\frac{3}{2}}-
\frac{1}{3}\frac{\Phi^{'}}{\Phi}\Omega_{\Lambda}-\frac{\lambda_{1}}{3}\Omega_{c}
-\frac{\lambda_{2}}{3}\Omega_{\Lambda}+\frac{\alpha}{3}(\Omega_{r}+\Omega_{m}+\Omega_{\Lambda})
]-\frac{2}{c}\Omega_{\Lambda}^{\frac{1}{2}}.\label{omegaLp}
\end{equation}
In addition, using $\Phi=\Phi_{0}(\frac{a}{a_{0}})^{\alpha}$,
$\Phi'=\frac{d \Phi}{d \ln a}=\alpha \Phi$, and
$\Omega_{r}+\Omega_{m}+\Omega_{\Lambda}=1$,   the derivative
expressions of Hubble parameter and energy density for TVCC as a
function of redshift $z$, can be
 expressed,
\begin{equation}
\frac{dH}{dz}=-\frac{H}{1+z}\frac{-9}{6+6\alpha-\omega\alpha^{2}}[\Omega_{r}+\Omega_{m}
+\frac{2}{3c}\Omega_{\Lambda}^{\frac{3}{2}}-\frac{\alpha}{3}\Omega_{\Lambda}
-\frac{\lambda_{1}}{3}\Omega_{c}-\frac{\lambda_{2}}{3}\Omega_{\Lambda}+\frac{\alpha}{3}],
\end{equation}
\begin{equation}
\frac{d\Omega_{\Lambda}}{dz}=-\frac{\Omega_{\Lambda}}{1+z}[\frac{18}{6+6\alpha-\omega
\alpha^{2}}(\Omega_{r}+\Omega_{m}+\frac{2}{3c}\Omega_{\Lambda}^{\frac{3}{2}}-\frac{\alpha}{3}
\Omega_{\Lambda}-\frac{\lambda_{1}}{3}\Omega_{c}
-\frac{\lambda_{2}}{3}\Omega_{\Lambda}+\frac{\alpha}{3})-\frac{2}{c}\Omega_{\Lambda}^{\frac{1}{2}}].
\end{equation}
And the deceleration parameter $q$ is derived,
\begin{equation}
q=-\frac{\ddot{a}}{aH^{2}}=\frac{2}{2+\alpha}[\frac{1}{2}+\Omega_{\Lambda}(\frac{1}{c}\Omega_{\Lambda}^{\frac{1}{2}}-\frac{1}{2}\alpha
-\frac{1}{2}\lambda_{1}\frac{\Omega_{c}}{\Omega_{\Lambda}}-\frac{1}{2}\lambda_{2}-\frac{3}{2})
+\alpha(\frac{\omega}{4}\alpha+\frac{1}{2}\alpha+\frac{1}{2})].\label{q}
\end{equation}
In above equations, for $\lambda_{1}=\lambda_{2}=0$, they reduce to
the non-interacting cases; for $\omega\rightarrow \infty$ and
$\alpha=0$, they correspond to the time variable cosmological
constant  in GR theory.

  %\begin{figure}[!htbp]
  %\includegraphics[width=4cm]{iage-bd-q-l.eps}
  %~~~\includegraphics[width=4cm]{iage-bd-q-l1.eps}
  %~~~\includegraphics[width=4cm]{iage-bd-q-l2.eps}\\
  %\caption{The evolution of $q(z)$ for IAge-BD.} \label{figurewde-qage}
  %\end{figure}

  %\begin{figure}[!htbp]
  %\includegraphics[width=4cm]{iage-bd-w-l.eps}
  %~~~\includegraphics[width=4cm]{iage-bd-w-l1.eps}
  % ~~~\includegraphics[width=4cm]{iage-bd-w-l2.eps}\\
  %\caption{The evolution of $w(z)$ for IAge-BD.} \label{figurewde-qage}
  %\end{figure}

 \begin{figure}[!htbp]
  \includegraphics[width=5cm]{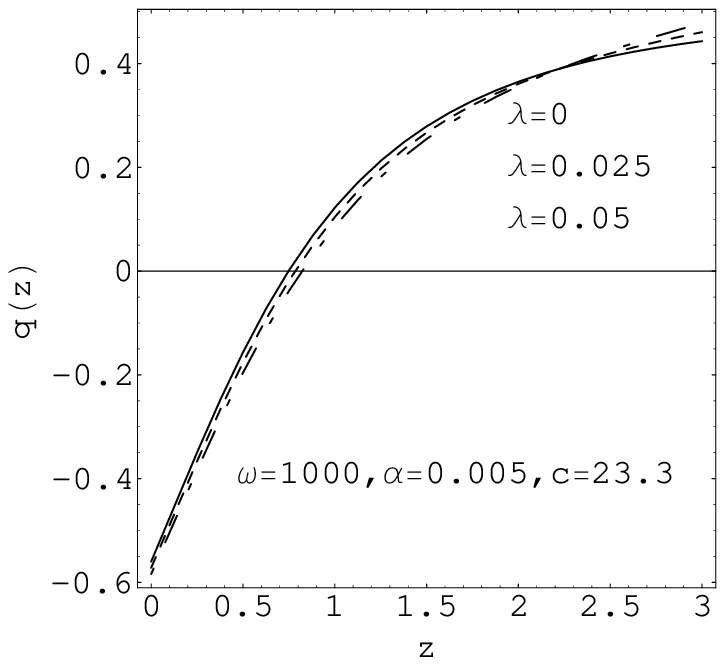}
   ~~~~~~\includegraphics[width=5cm]{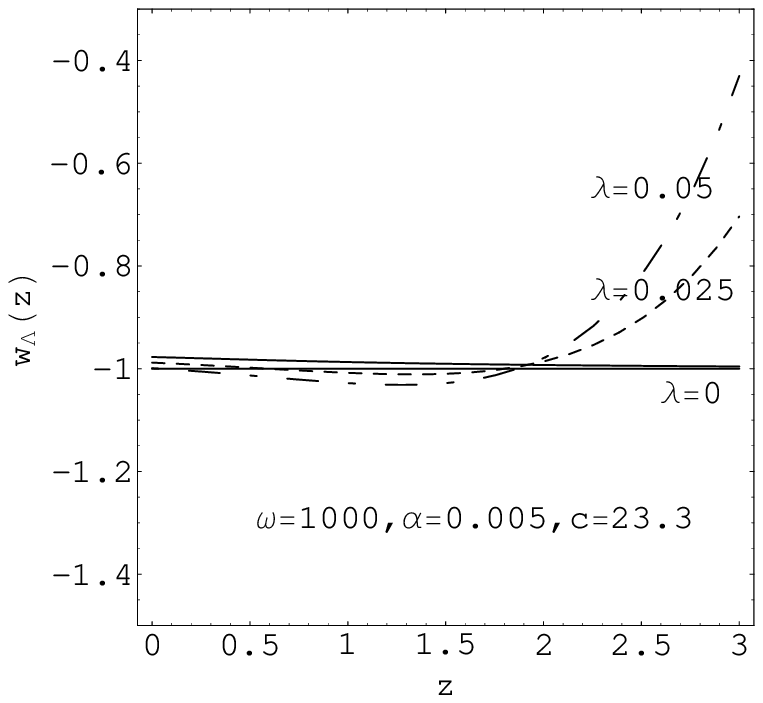}\\
  \caption{The evolutions of EOS  $w_{\Lambda}(z)$ and decelerating parameter  $q(z)$
    for interacting time variable cosmological constant in Brans-Dicke theory. } \label{figure-iage-bd-qw}
\end{figure}

 According to the Eqs.(\ref{w}) and  (\ref{q}), the evolutions of EOS
 $w_{\Lambda}(z)$ and decelerating parameter  $q(z)$
 for interacting time variable cosmological constant in Brans-Dicke theory are illustrated in Fig.
 \ref{figure-iage-bd-qw}, with the values of model parameters $\omega=1000$ \cite{w-1000},
 $\alpha=0.005$ \cite{Lu-bd}, $c=23.3$ \cite{Chen-c}, where
 the three lines correspond respectively to the cases of parameter
 $\lambda=0$, $\lambda=0.025$, $\lambda=0.05$
 for interacting case $Q=\lambda H(\rho_{c}+\rho_{\Lambda})$. According to the Fig. \ref{figure-iage-bd-qw},
 it can be seen that an expansion of
 universe from deceleration to acceleration is given in this model.
 And one can see that the shape of deceleration parameter $q(z)$ is
 not sensitive to the values of interaction parameter $\lambda$. For
 the evolution of EOS $w_{\Lambda}(z)$, at low redshift it is similar
 to cosmological constant, and  at high redshift its evolution  is
 dependent on the values of interaction parameter. In addition, in
 this scenario it is shown that the evolution of EOS $w_{\Lambda}(z)$
 can cross over the boundary of phantom divide \cite{cross}, as
 realized
 in the quintom model \cite{quintom}. For the case of
 $\lambda=0.025$, the values of transition redshift, current
 deceleration parameter and EOS are given respectively as
 $z_{T}=0.784$, $q_{0}=-0.572$, and $w_{0\Lambda}=-0.988$,  which are
 consistent with the current observational constraint, where the
 model independent scenarios are constrained from the latest
 observational data \cite{lu-mpla}.

\begin{figure}[!htbp]
  \includegraphics[width=7cm]{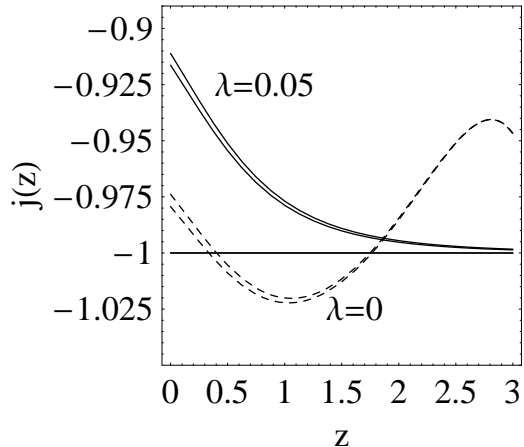}\\
  \caption{The evolutions of $j(z)$ for  time variable cosmological constant in Brans-Dicke theory and
  in general relativity, respectively.} \label{figure-iage-bd-j}
\end{figure}

Next we use a geometrical diagnostic method, i.e. jerk parameter,
to distinguish the cosmological constant model and the interacting
TVCC. Jerk parameter $j$ is defined as,
\begin{equation}
j \equiv
-\frac{1}{H^{3}}(\frac{\dot{\ddot{a}}}{a})=-[\frac{1}{2}(1+z)^{2}\frac{[H(z)^{2}]^{''}}{H(z)^{2}}
-(1+z)\frac{[H(z)^{2}]^{'}}{H(z)^{2}}+1].\label{jerk}
\end{equation}
 With the help of parameter $j$, the transitions between
phases of different cosmic acceleration can be more  conveniently
described. And Eq. (\ref{jerk}) can be derived as
\begin{equation}
 j=-1-\frac{9}{2}w_{de}\Omega_{de}(1+w_{de})+\frac{3}{2}\Omega_{de}\frac{\dot{w}_{de}}{H},\label{jlcdm}
 \end{equation}
 where $\Omega_{de}$ denotes dimensionless energy density
 for dark energy, and EOS for dark energy $w_{de}$  can be either larger or smaller than -1.
  Also, from Eq. (\ref{jlcdm}) it is easy to see that,
  for the cosmological constant model ($w_{\Lambda}(z)=-1$), it has a constant jerk
 with $j(z) = -1$. Thus  the jerk
 parameter can provide us with a  convenient approach  to
 search for departures  from the cosmic concordance model,
  just as deviations from $w_{\Lambda}(z) = -1$ done in more
 standard dynamical analysis.
According to Eq.(\ref{jlcdm}), the evolutions of jerk parameter
$j(z)$ are plotted in Fig. \ref{figure-iage-bd-j} for different
cases, where the two dotted lines denote the cases of
non-interacting time variable cosmological constant in BD and GR
gravitational theory,  respectively, as well as  two solid lines
denoting interacting cases. It is shown that the evolutions of
$j(z)$ are similar in these two gravitational theories for both
   interacting and non-interacting TVCC. Also, in Fig. \ref{figure-iage-bd-j} one can see the
departures  from the cosmological constant for these four cases.

\section{$\text{Time variable cosmological constant in Brans-Dicke theory}$}

 \begin{figure}[ht]
    \includegraphics[width=5cm]{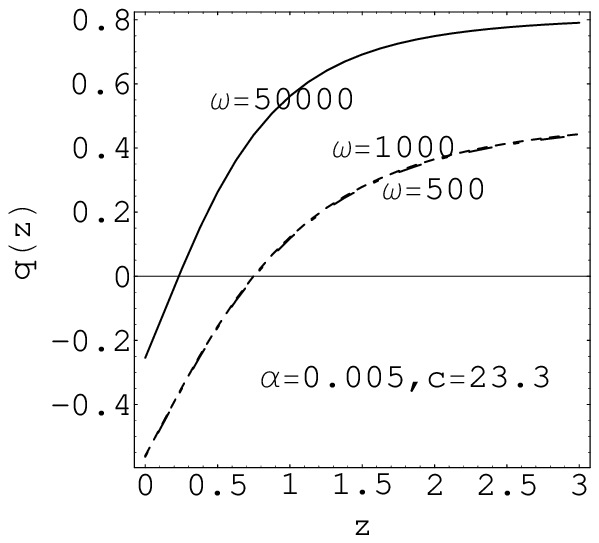}
 ~~~\includegraphics[width=5cm]{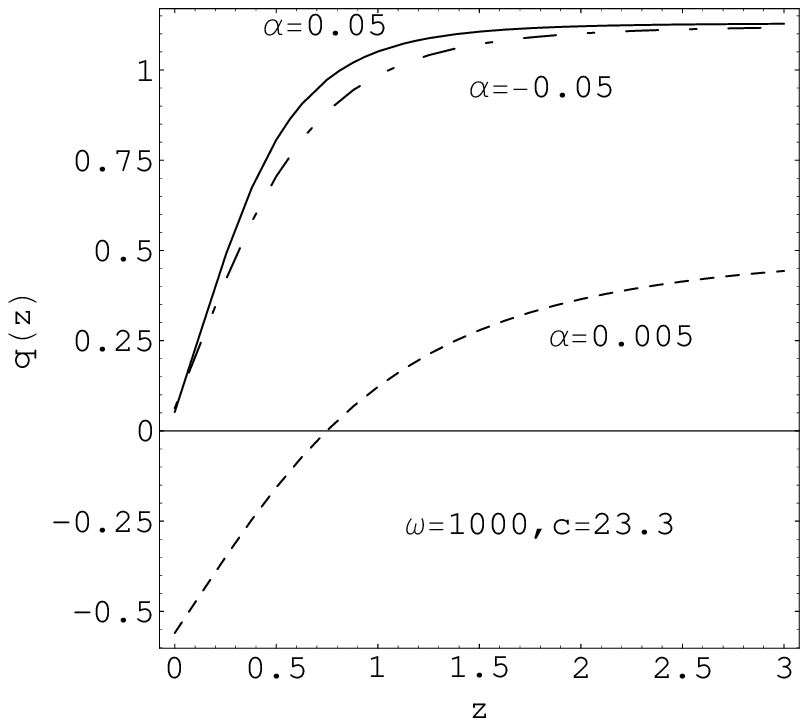}
 ~~~\includegraphics[width=5cm]{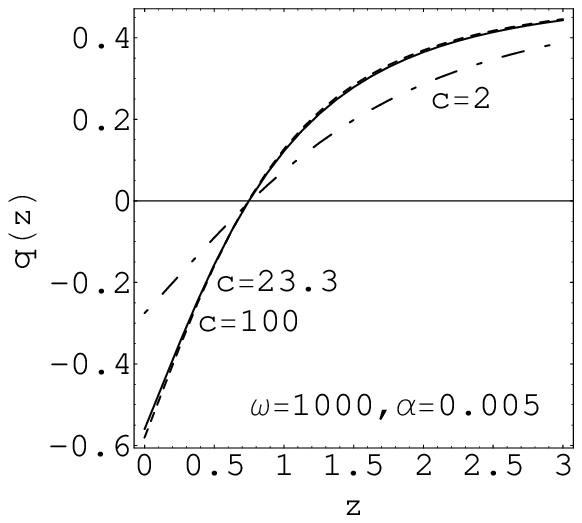}\\
  \caption{The evolutions of $q(z)$ for time variable cosmological constant in Brans-Dicke theory
   with using various model parameters.} \label{figure-age-bd-q}
 \end{figure}
 \begin{figure}[ht]
   \includegraphics[width=5cm]{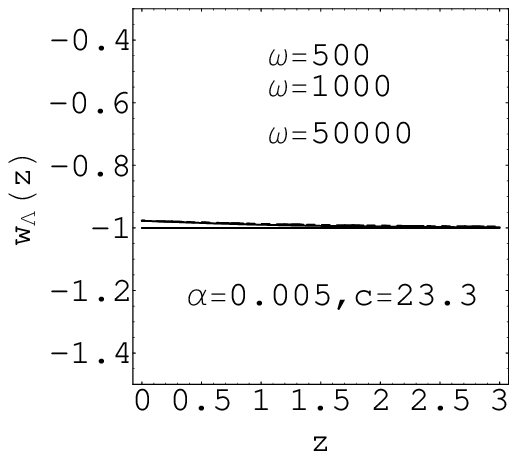}
 ~~~\includegraphics[width=5cm]{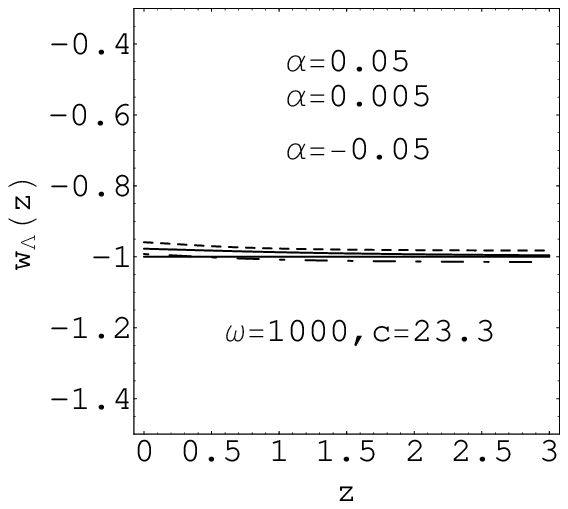}
 ~~~\includegraphics[width=5cm]{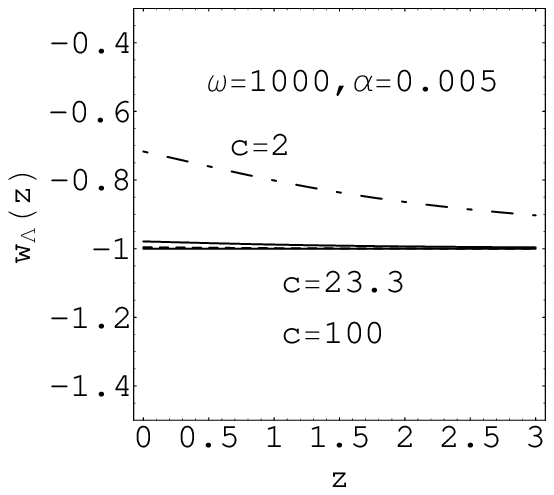}\\
  \caption{The evolutions of $w_{\Lambda}(z)$ for time variable cosmological constant in Brans-Dicke theory
   with using various model parameters.}\label{figure-age-bd-w}
 \end{figure}

\begin{figure}[!htbp]
  ~~~\includegraphics[width=5cm]{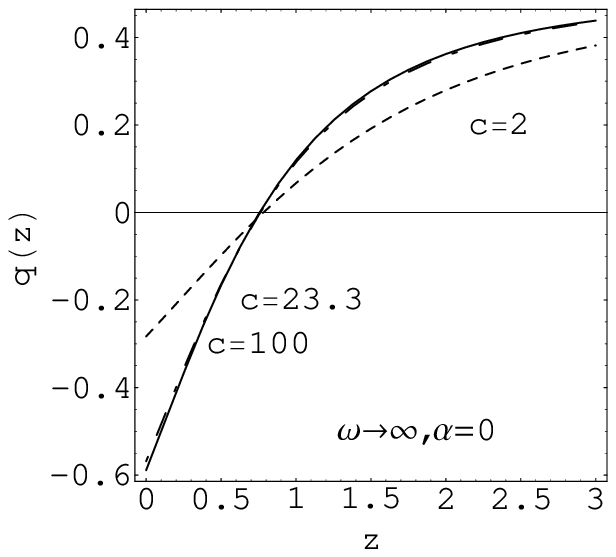}
 ~~~\includegraphics[width=5cm]{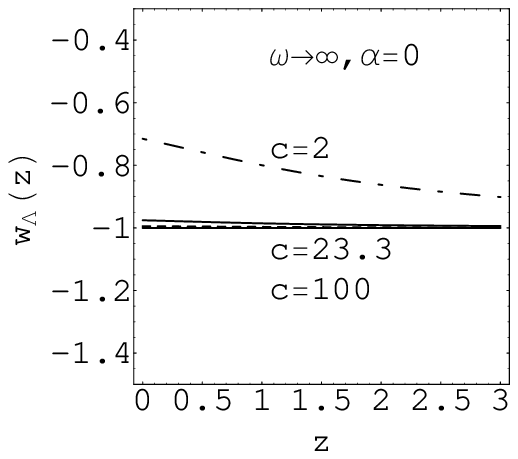}\\
  \caption{The evolutions of $q(z)$ and $w_{\Lambda}(z)$ for time variable cosmological
constant  in GR theory
   with using different values of $c$.}\label{figure-age-nobd}
 \end{figure}

Next we investigate the evolutions of cosmological quantities for
taking different values of model parameters to show their dependence
on model parameters.  We firstly consider the case of TVCC with no
interaction in Brans-Dicke theory. Here the evolutions of
deceleration parameter and EOS for TVCC model are plotted in Figs.
\ref{figure-age-bd-q} and \ref{figure-age-bd-w}, with various
parameters: (1) $\omega$=(500, 1000, 50000), $\alpha$=0.005, c=23.3;
(2) $\omega=1000$, $\alpha$=(-0.05, 0.005, 0.05), c=23.3; (3)
$\omega=1000$, $\alpha=0.005$, c=(2, 23.3, 100). From the evolutions
of $q(z)$ and $w_{\Lambda}(z)$ in Figs. \ref{figure-age-bd-q} and
\ref{figure-age-bd-w}, one can see that $q(z)$ is more dependent on
the values of $\alpha$ and $\omega$, but $w_{\Lambda}$(z) is
relative less dependent on them.  And for the case of
$\omega\rightarrow \infty$, $\alpha=0$ in the framework of GR
theory, $q(z)$ and $w_{\Lambda}$(z) are plotted in Fig.
\ref{figure-age-nobd} with  c=(2, 23.3, 100). Comparing this figure
with Figs. \ref{figure-age-bd-q} and \ref{figure-age-bd-w}, we can
see that the differences of the evolutions of $q(z)$ and
$w_{\Lambda}(z)$ between BD theory and GR theory are not obvious.

\section{$\text{Time variable  cosmological constant with interaction}$}

\begin{figure}[!htbp]
   \includegraphics[width=5cm]{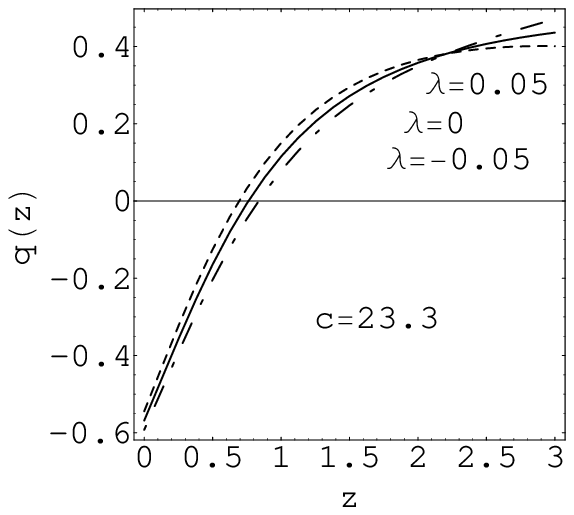}
 ~~\includegraphics[width=5cm]{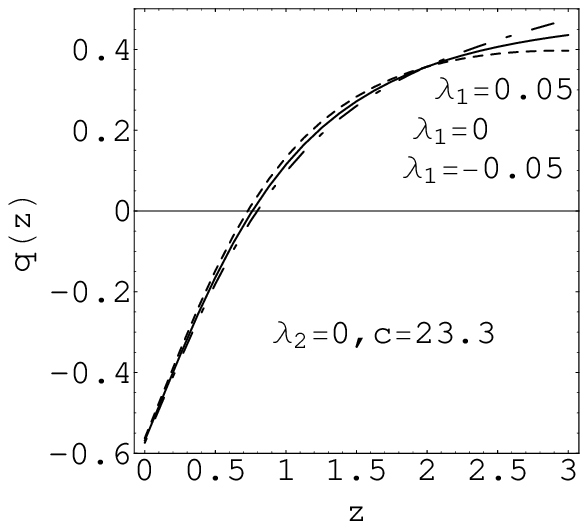}
 ~~\includegraphics[width=5cm]{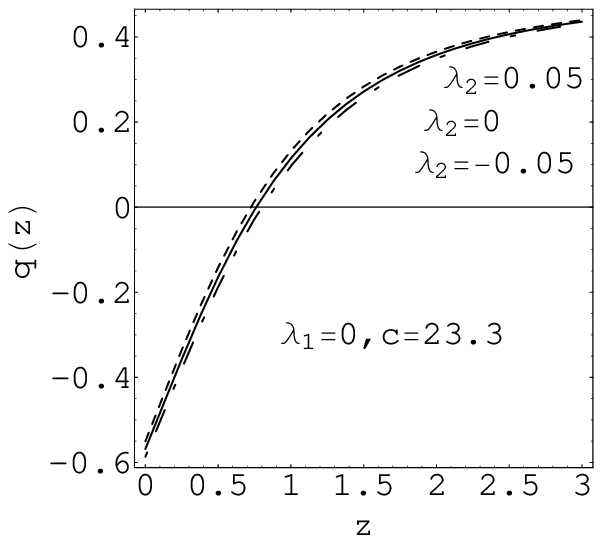}\\
  \caption{The evolutions of $q(z)$ for interacting time variable cosmological constant.} \label{figure-iage-q}
\end{figure}

\begin{figure}[ht]
  \includegraphics[width=5cm]{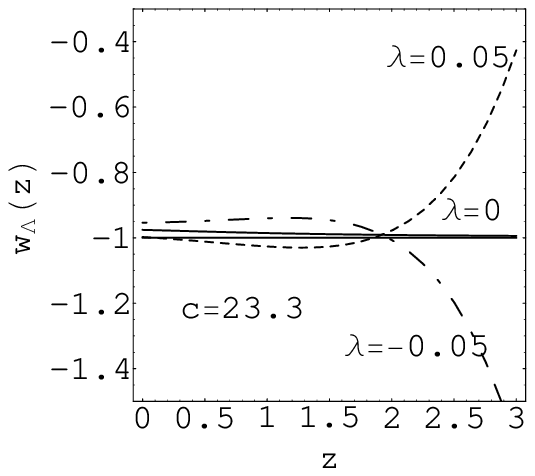}
 ~~~\includegraphics[width=5cm]{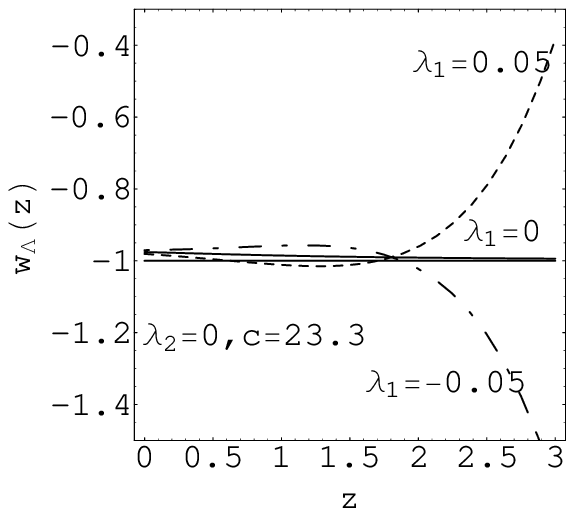}
 ~~~\includegraphics[width=5cm]{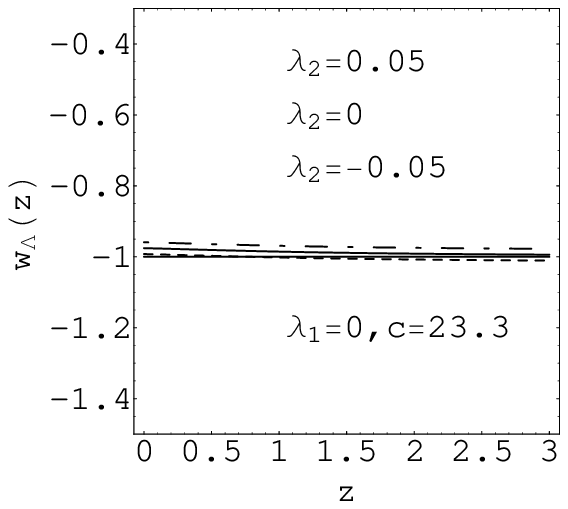}\\
  \caption{The evolution of $w_{\Lambda}(z)$ for interacting time variable cosmological constant.} \label{figure-iage-w}
\end{figure}

\begin{figure}[ht]
 ~~\includegraphics[width=5cm]{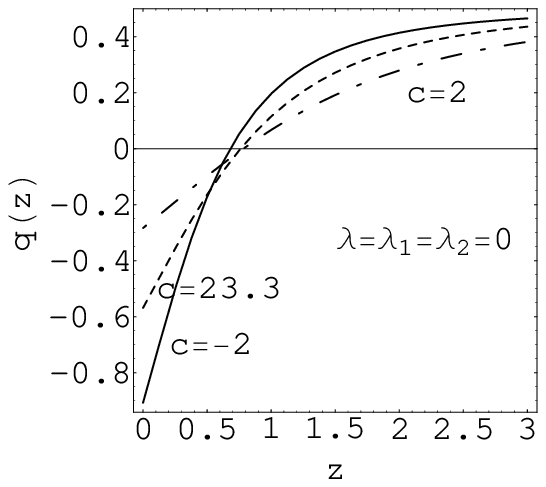}
 ~~~\includegraphics[width=5cm]{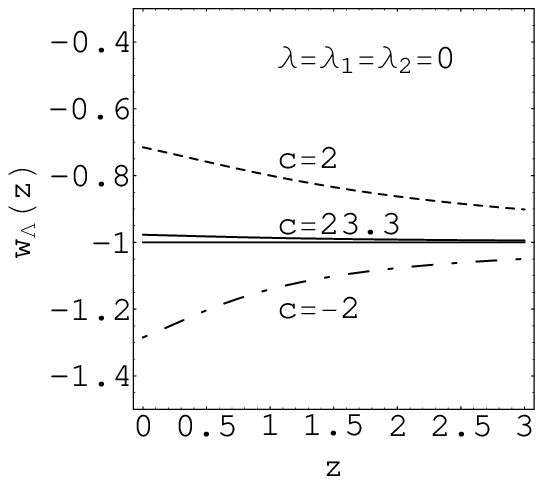}\\
  \caption{The evolutions of $q(z)$ and $w_{\Lambda}(z)$ for
   non-interacting time variable cosmological constant.} \label{figure-age}
\end{figure}

In this part we consider different interactions for time variable
cosmological constant in GR theory, i.e. taking $\omega\rightarrow
\infty, \alpha=0$. For  Figs. \ref{figure-iage-q} and
\ref{figure-iage-w}, the parameters are taken respectively  as, (1)
$c=23.3$, $\lambda=\lambda_{1}=\lambda_{2}$=(0.05, 0, -0.05); (2)
 $c=23.3$, $\lambda_{1}$=(0.05, 0, -0.05), $\lambda_{2}=0$;
 (3)  $c=23.3$,  $\lambda_{1}=0$, $\lambda_{2}$=(0.05, 0, -0.05).
According to the Figs.  \ref{figure-iage-q} and \ref{figure-iage-w}
one can see that, the shape of deceleration parameter $q(z)$ is not
sensitive to the
 values of interaction parameter and its interacting forms;
for the evolution of $w_{\Lambda}(z)$ it is not sensitive to the
variable of parameter values $\lambda_{2}$, and they are  similar
for the cases of $Q=\lambda H(\rho_{c}+\rho_{\Lambda})$ and
$Q=\lambda_{1}H\rho_{c}$. And from Fig.  \ref{figure-iage-w}, it can
be seen that the EOS can cross over the phantom boundary
$w_{\Lambda}=-1$ for the exist of the interacting parameter
$\lambda$ or $\lambda_{1}$. But for the case of negative interaction
parameter $\lambda<0$ (or $\lambda_{1}<0$), it is found that there
is a smaller value of EOS at higher redshift. In addition, in Fig.
\ref{figure-age} we show the
 non-interacting case $\lambda=\lambda_{1}=\lambda_{2}=0$ with using $c$=(-2, 2, 23.3).
According to  Fig. \ref{figure-age}, it is indicated that for using
the parameter value $c\in(2,\infty)$ and $c\in(-\infty,-2)$, the
current value of $w_{0\Lambda}$ lies in (-1.285, -0.715), and the
value of current deceleration parameter $q_{0}\in $ (-0.907,
-0.283), transition redshift $z_{T}\in$ (0.681, 0.782).

\section{$\text{Conclusions}$}

In this paper the time variable cosmological constant with
interaction in Brans-Dicke theory is discussed. One can see that an
accelerating universe at late time is obtained in this model. And
the equation of state for the interacting case can cross over the
phantom boundary.  Furthermore, by taking the model parameters to be
in  large regions,  $\omega \geq500$, $\alpha \in[-0.05,0.05]$,
$\lambda \in[-0.05,0.05]$, we investigate the evolutions of
cosmological quantities for time variable cosmological constant. It
is shown that the shape of deceleration parameter $q(z)$ is not
sensitive to the
 values of interaction parameter and its interacting forms, but it is
relatively more dependent on the values of BD parameter $\alpha$ and
$\omega$. For the evolution of EOS $w_{\Lambda}(z)$, it is more
dependent on the values of interaction parameter, but is less
sensitive to the variation of BD parameter $\alpha$ and $\omega$. In
addition we can see that for the evolutions of deceleration
parameter, EOS, and jerk parameter geometrical method, the
differences between BD and GR theory are not obvious  for time
variable cosmological constant.

 \textbf{\ Acknowledgments }
 The research work is supported by the National Natural Science Foundation
  (Grant No. 10875056), NSF (10703001) and NSF (No.11005088) of P.R. China.

\appendix
\section{$\text{Investigate interaction between dark sections with a different
 interacting form $Q=\lambda_{3}\dot{\rho}_{\Lambda}$}$}\label{appendix}

\begin{figure}[ht]
 ~~\includegraphics[width=5cm]{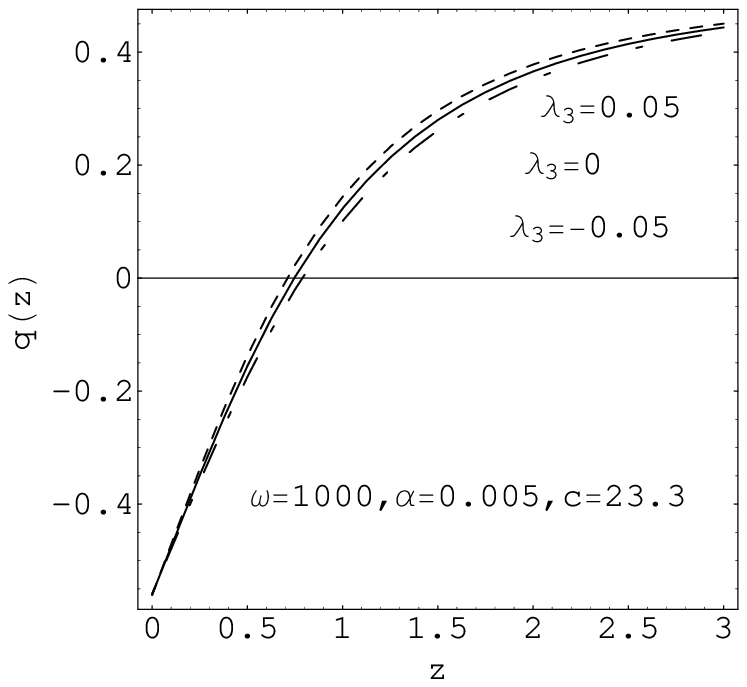}
 ~~~\includegraphics[width=5cm]{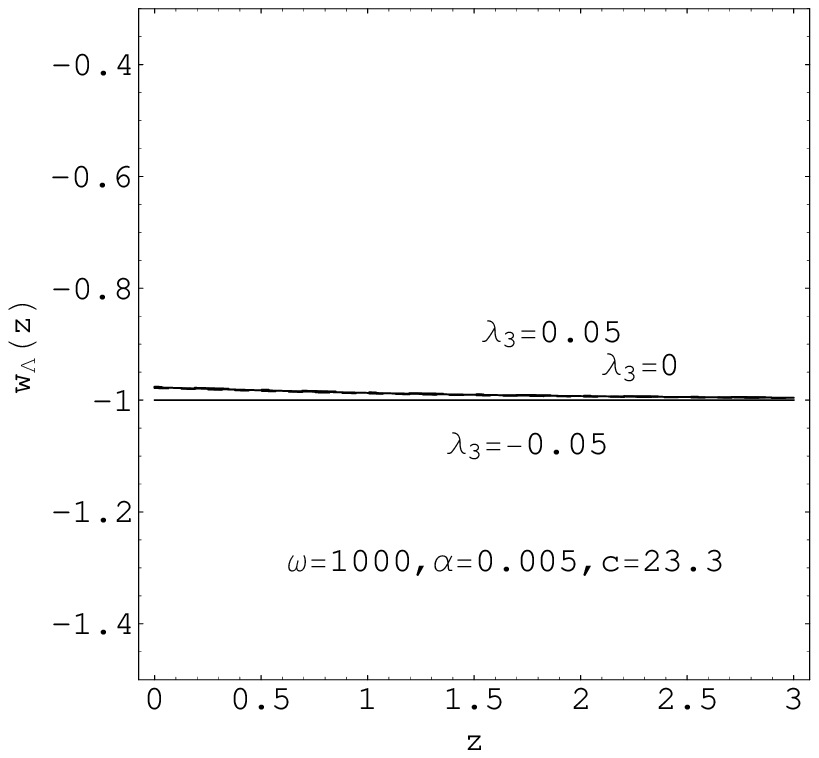}\\
  \caption{The evolutions of $q(z)$ and $w_{\Lambda}(z)$ for
   interacting time variable cosmological constant in Brans-Dicke theory, with the interaction from
    $Q=\lambda_{3}\dot{\rho}_{\Lambda}$.} \label{qw-l3}
\end{figure}

Also, for comparison we consider another interacting form $Q=
\lambda_3 \dot{\rho}_{\Lambda}$, which is consistent with the
analysis of dimension according to the Eq. (\ref{IL}). Obviously,
this interacting term is different from the above one $Q= \lambda_1
H \rho_c+\lambda_2 H \rho_{\Lambda}$. For this case the
corresponding  expressions of some cosmological quantities in BD
theory are derived as
\begin{equation}
w_{\Lambda}=\frac{2}{3c}\Omega_{\Lambda}^{\frac{1}{2}}+\frac{2\lambda_{3}}{3c}\Omega_{\Lambda}^{\frac{1}{2}}
-\frac{\alpha}{3}-\frac{\alpha \lambda_{3}}{3}-1,\label{wl3}
\end{equation}
\begin{equation}
\frac{d\Omega_{\Lambda}}{dz}=-\frac{\Omega_{\Lambda}}{1+z}[\frac{18}{6+6\alpha-\omega
\alpha^{2}}(\Omega_{r}+\Omega_{m}+\frac{2}{3c}\Omega_{\Lambda}^{\frac{3}{2}}
+\frac{2\lambda_{3}}{3c}\Omega_{\Lambda}^{\frac{3}{2}}-\frac{\alpha}{3}
\Omega_{\Lambda}-\frac{\alpha \lambda_{3}}{3}
\Omega_{\Lambda}+\frac{\alpha}{3})-\frac{2}{c}\Omega_{\Lambda}^{\frac{1}{2}}
-\frac{2\lambda_{3}}{c}\Omega_{\Lambda}^{\frac{1}{2}}+\lambda_{3}\alpha-\lambda_{3}],\label{Omegal3}
\end{equation}
\begin{equation}
q=\frac{2}{2+\alpha}[\frac{1}{2}+\Omega_{\Lambda}
(\frac{1}{c}\Omega_{\Lambda}^{\frac{1}{2}}+\frac{\lambda_{3}}{c}\Omega_{\Lambda}^{\frac{1}{2}}-\frac{1}{2}\alpha
-\frac{\lambda_{3}}{2}\alpha-\frac{3}{2})
+\alpha(\frac{\omega}{4}\alpha+\frac{1}{2}\alpha+\frac{1}{2})],\label{ql3}
\end{equation}
with using a parameterized form for BD scalar field
$\Phi=\Phi_{0}(\frac{a}{a_{0}})^{\alpha}$. And the evolutions of
deceleration parameter and EOS for this interacting time variable
cosmological constant in the framework of BD theory are plotted in
Fig.\ref{qw-l3}. From this figure we can see that, comparing with
the non-interacting case the influences on the evolutions of $q(z)$
and $w_{\Lambda}(z)$ from this interacting form are not obvious, so
we do not discuss  much for this case in the text.

\end{document}